\documentclass[aps,prl,twocolumn,superscriptaddress,groupedaddress]{revtex4}
\usepackage{subfig}
\usepackage{graphicx}  
\usepackage{dcolumn}   
\usepackage{bm}        
\usepackage{amssymb}   
\usepackage{slashed}
\usepackage{graphicx}				
\usepackage{amsmath}
\usepackage{mathtools}
\usepackage{tikz,pgf}
\usepackage{comment}
\usepackage{asymptote}
\usetikzlibrary{arrows,backgrounds}
\usetikzlibrary{fit,scopes,calc,matrix,positioning,decorations.pathmorphing}
\usepackage[all]{xy}
\usepackage{yfonts}

\newcommand{\ket}[1]{\ensuremath{\left|#1\right\rangle}}
\newcommand{\Bracket}[1]{\ensuremath{\left\langle#1\right\rangle}}

\begin{document}
\title{Entanglement as a resource for naturalness}
\author{Andrei T. Patrascu}
\address{ELI-NP, Horia Hulubei National Institute for R\&D in Physics and Nuclear Engineering, 30 Reactorului St, Bucharest-Magurele, 077125, Romania}
\begin{abstract}
A novel approach to understanding the hierarchy problem is presented making use of topological aspects of the renormalisation group and the ER-EPR interpretation of entanglement. A common discussion of the renormalisation group, the black hole horizon, and the expected entanglement between outgoing Hawking radiation and the interior, the cosmic censorship mechanism and the cosmological constant problem is envisaged. 
\end{abstract}
\maketitle
\section{Introduction}
The discovery of the Higgs boson at the frontier between the regions of stability and metastability of the standard model vacuum with no other supersymmetric partners found at reasonable energies has triggered many questions regarding naturalness. As it stands, the standard model seems un-natural. Of course this state of affairs is only temporary and totally unsatisfactory from the perspective of understanding Nature. The Higgs boson is the only fundamental scalar of the standard model (known so far) and it is expected that radiative corrections in the absence of a protecting supersymmetry would bring its mass up to the Planck scale. Understanding why the Higgs mass still remains at relatively low energies, comparable to the electroweak scale is regarded as a mystery. One can turn this problem around and realise that the Higgs scalar may represent a relevant degree of freedom which is strongly dependent on the physics at and beyond the UV cut-off of our theory. Its mass certainly is strongly influenced by such high energy degrees of freedom and hence a protection mechanism should be in place so that its mass becomes natural at the low energies where it has been observed. Overall, the question translates into how we can explain away unnatural phenomena considering the idea that otherwise the Standard Model is up to a very high degree of accuracy an effective field theory having its underlying degrees of freedom well hidden. We know of several mechanisms through which the high energy degrees of freedom are being hidden from a low energy observer. In gravity, high curvature physics is being screened via a "cosmic censorship" principle realised in the form of a black hole horizon. In cosmology the degrees of freedom at the origin of the universe are being washed away by the process of inflation, while finally, in effective quantum field theories, the high energy degrees of freedom are being hidden by Universality, a property of the renormalisation group which says that only a very small number of directions in the parameter space would bring us away from a fixed point, while the others would pull us back. This basically means that many high energy theories would lead to the same macroscopic low energy theory, which would make it particularly hard for us to discern high energy degrees of freedom from physics at our scale. All these mechanisms of hiding away high energy degrees of freedom however have loopholes. In the case of black holes, the Hawking radiation is a phenomenon that occurs due to restrictions imposed on the quantum field modes in the spacetime around a black hole. Such restrictions make the requirement of an outgoing radiation evident. It is assumed that entanglement (equivalently wormhole geometry or instanton effects, basically topologically non-trivial effects) would convey the information behind the horizon to an outside observer through various correlations. A similar process occurs in the case of the Unruh effect and the associated horizon together with the particles detected by an accelerating observer in vacuum. In the case of cosmology, early-late universe entanglement is considered to be manifest and could play an important role in understanding the problem of the cosmological constant. 

Together with this, I showed [1] that T-duality in string theory may hold a fundamental meaning in the description of inflation. I also showed that topologically non-trivial effects shifting the mass of the axion can be interpreted as quantum entanglement [7]. In the case of the renormalisation group and for the Higgs mass a similar effect may be at work. While performing a renormalisation group calculation, the basic idea is to integrate (trace) over the high energy modes and to implement the effects on the low energy physics just by allowing the flows of the coupling constants. This mechanism is generally possible while not reversible (hence the renormalisation group is actually just a semi-group). What could be the loophole in this situation? 
In general, the renormalisation group approach is seen as a form of simplification, or a reduction of the complexity associated to the microscopic degrees of freedom by means of replacing them with macroscopic (or lower energy) effective degrees of freedom. This type of thinking led to a worldview based on so called "reductionism" in which successive simplifications and reductions in complexity can be performed when transitioning to lower energy / higher length scales. 
I believe this worldview is misleading, placing too much focus on the idea of complexity and ignoring the fallibility of our methods of describing Nature. In fact, there is no reason to believe that the complexity of the underlying degrees of freedom is somehow reduced, and instead, it could be that the more our methods and formal languages become compatible with the natural ones, the easier the calculations can be performed, not due to a decrease in complexity, but due to a compatibility of formal languages. To be more specific here, in a renormalisation group approach we do indeed first integrate out the lower scale degrees of freedom and replace them to more appropriate degrees of freedom for the scale we are working, while being very well aware that we ignore a series of deep UV effects in the process. We just hope that the processes that we ignore will not affect our ability to answer certain questions. For the most part they do not. However, for some situations, as is the case of massive scalars, they do. Therefore, it appears that it was not the reduction of the number of degrees of freedom, and therefore of the complexity of the problem, that was necessary, but instead the adoption of more suitable degrees of freedom, and all in all, of a more suitable formal language for the given scale. To describe the physics of massive scalars however, we do need a formal language that has a global structure that is more compatible with Nature and probably with correlations that are washed away by our renormalisation techniques. Therefore, it is my worldview that it is not the reduction in complexity that made the renormalisation group so desirable, but the replacement of one formal language, written in terms of unphysical or poorly guessed parameters and degrees of freedom, with one that has more suitable degrees of freedom at a given scale. Doing the same thing in a global sense for all scales is still a work in progress. 

We know that at the level of the cut-off scale we must stop trusting our theory as new degrees of freedom may have significant effects. Restricting ourselves to only low energy modes of our field theory somehow reminds us of the process of restricting modes in the case of the Hawking radiation. One could imagine a way in which this cut-off could be assimilated to a form of "horizon" for our theory with nothing beyond it affecting the theory above it, except for an interesting flow of the parameters of the theory. Still, there will be loopholes here too. Topologically non-trivial phenomena may have a non-trivial impact on the parameters of the low energy theory in the same way in which entanglement (seen as a non-trivial spacetime topology via the ER-EPR duality) plays a role in the Hawking radiation emitted from the vicinity of the black hole. Topological effects in the Renormalisation group have been studied in the context of T-duality in ref. [1]. There I showed that demanding invariance from changes of topology amounts to corrections to the Higgs mass that could restore naturalness. Here, I expand this idea considering a broader interpretation of a horizon triggered entanglement and show that such topological terms (as entanglement is seen as a topological term, see ref. [2] and [3]) will bring us information beyond the scales to which we would otherwise have access. This article will have 4 parts. In the next section I will briefly present the hierarchy problem from the perspective of a coupling between scales showing that the lack of naturalness can be related to the idea that scales are not separated in precisely the way an effective quantum field theory would tell us. This discussion has basically been done also in [4] which I use as a reference. In the next section I will start with a simple approach to a renormalisation group problem and will show what happens if topological terms are present. I will show what effects are to be expected in the case of scalar fields and their masses. In the next section I will discuss this mechanism in the context of the analogies with the cosmic censorship, Hawking radiation, and the cosmological constant, arguing that the solution to all these problems is of the same type: topologically non-trivial phenomena connecting apparently distinct scales and restricting the available modes in quantum field theory. The obvious connection to string-theoretical T-duality and its way of connecting all scales is briefly discussed. Finally, some basic conclusions are being outlined.
\section{Naturalness and relevant degrees of freedom}
Nature seems to be hiding things from us, using mechanisms from the same category. High energy degrees of freedom within a black hole are being hidden by a classical horizon. General relativity seems to tell us that nothing passing that horizon can ever come back or transmit any information outside. Still, quantum field theoretical and finally topological phenomena (modes restrictions via the Hawking mechanism and entanglement seen as a topological effect by means of ER-EPR) tell us that we may have some hope of detecting effects of such hidden degrees of freedom after all. Inflation seems to hide the initial degrees of freedom of our universe by a similar mechanism which may be overcome by thinking in terms of extraction of entanglement from the vacuum. Finally, the universality of the renormalisation group approach seems to wash away any degrees of freedom reminding us of the high energy physics by means of a "critical surface". There are only very few types of terms to be added to the functional form of our action that are relevant, hence could bring us out and affect the low energy physics. These terms can be seen as topologically non-trivial in the parameter space and hence expressible in terms of entanglement via ER-EPR. 
In ref. [4] a simple model is employed to show how naturalness fails in the case of renormalisation group flows of unnatural couplings. Their example is a simple theory with a scalar field $\phi$ with bare mass $m$ and a massive fermion field $\psi$ with a mass $M$ interacting by means of a Yukawa interaction with coupling $g$. The theory can be written as 
\begin{equation}
\mathcal{L}=\frac{1}{2} \partial_{\nu}\phi\partial^{\nu}\phi-\frac{1}{2}m^{2}\phi^{2}-\frac{\lambda}{4!}\phi^{4}+i\bar{\Psi}\gamma^{\nu}\partial_{\nu}\Psi - M \bar{\Psi}\Psi+g\phi\bar{\Psi}\Psi
\end{equation}
We may consider first the scalar mass much higher than the fermion mass $m >> M$ and we study the physics at some low energy scale $E<<M$. We can use perturbation theory and calculate the effects of the heavy scalar field in the full theory and then integrate it out incorporating its effects in the couplings of the light fermion field in the low energy theory. The resulting effective theory will only include the light fermion field with modified (shifted) couplings $M^{*}$ and $g^{*}$. The resulting effective theory will be [4]
\begin{equation}
\mathcal{L}=i\bar{\Psi}\gamma^{\nu}\partial_{\nu}\Psi-M^{*}\bar{\Psi}\Psi + \frac{g^{*}}{2m^{2}}(\bar{\Psi}\Psi)^{2}
\end{equation}
In general the effective mass will be equal to the bare mass plus corrections coming from the high energy physics. If we calculate them up to one loop they look like 
\begin{equation}
M^{*}=[M+M\frac{g}{16 \pi^{2}}ln(\frac{\Lambda}{M})]=(M+\Delta M)
\end{equation}
The correction to the bare mass is proportional to the bare mass itself which, considering it was small in the original theory, will remain small in the effective theory as well. If however the situation changes and the fermion mass is much heavier than the scalar mass we will have a very different result
\begin{equation}
(m^{*})^{2} = m^{2}+\frac{g}{16 \pi^{2}}[\Lambda^{2} + M^{2}+m^{2}ln(\frac{\Lambda}{M})+\mathcal{O}(\frac{M^{4}}{\Lambda^{4}})]
\end{equation}
This shows that starting with a small scalar mass perturbative corrections can produce corrections of the magnitude of the cut-off scale $\Lambda$ where out theory would break down anyways. The procedure of integrating away the high energy fields did not work as planned. The resulting theory would contain a field with a mass proportional to the cut-off. One could imagine the scalar to be the Higgs field and the heavy fermion ot be the top quark. We expect the cut-off scale of the standard model to be at most the Planck scale and at least $1$ TeV (following [4]). However, even a correction of the order of ($1$ TeV)$^{2}$ would push the mass of the Higgs up by six orders of magnitude. From the perspective of the renormalisation group this would show how relevant and irrelevant parameters at low energy depend on their initial high energy values at or beyond the cut-off scale. Indeed, in the case of the Higgs the coupling and mass at the low energy end would depend strongly on extremely small changes in the cut-off couplings. Such a strong sensitivity is a benchmark for relevant degrees of freedom and for the Hierarchy problem. Therefore our cut-off region, aside of looking analogue to a horizon could also be regarded as the equivalent of a "fast scrambler" using quantum gravity terminology. If we look at the trajectories of the RG flows associated to the couplings/masses of the Higgs compared to other quantities, we see that them being relevant is translated into a topologically non-trivial link between the region beyond the cut-off and the region below. Such a topological feature reminds us of wormhole solutions and entanglement. Let us see in a simple example how a topologically non-trivial parameter space structure can play a significant role
\section{topology, entanglement, and the hierarchy problem}
The renormalisation group approach depends upon the fact that integrating over high momentum modes in the theory and shifting the scale of the theory results in an action that resembles the structure of the original action with the only distinction arising in the functional form of the coupling parameters (which start flowing). However, what would happen if additional topological structure is being added by the process of integrating over the modes? The high energy modes may have topologically non-trivial connections with the low energy ones. After all, if we think in terms of fixed points and scale independence at such fixed points, we come to the conclusion that fluctuations that we wash away by integration are being recovered from the next scale we consider in our calculations. Such fluctuations can be strongly entangled. It would be interesting to analyse this situation in the context of entanglement extraction from an intermediate vacuum state with early and late conformal backgrounds. That such states can be used for entanglement farming has been shown in [5] and [6]. Demanding invariance to such new topological changes would imply new terms to be visible at the lower energy domains. Such terms would offer non-trivial quantum correlation with the high energy physics and would appear as "wormhole corrections" in the low energy physics. Indeed, such "wormhole corrections" have recently been shown to shift the mass of the axion [7].
But let us see how this works. We can write our partition function as 
\begin{equation}
Z=\int \mathcal{D}\phi e^{-S[\phi]}
\end{equation}
let us establish a cut-off $\Lambda$ such that the Fourier modes of our fields vanish far above this cut-off
\begin{equation}
\phi_{k}=0, k>\Lambda
\end{equation}
As the renormalisation group procedure goes, we are only interested about physics at long length scales $L$ so we do not care about the modes $\phi_k>>1/L$. We may write our theory using a lower cut-off $\Lambda'=\frac{\Lambda}{\zeta}$ which is valid for $\Lambda'>>1/L$. Let me now write the Fourier modes as
\begin{equation}
\phi_{k}=\phi_{k}^{-}+\phi_{k}^{+}
\end{equation}
where $+$ describes the high energy modes.
Our action can be decomposed in terms of these modes as 
\begin{equation}
S[\phi_{k}]=S_{0}[\phi_{k}^{-}]+S_{0}[\phi_{k}^{+}]+S_{I}[\phi_{k}^{-}, \phi_{k}^{+}]
\end{equation}
Here the term $S_{0}[\phi_{k}^{-}, \phi_{k}^{+}]$ involves interactions between the high and low energy modes and hence a form of scale mixing. We can re-write our partition function as 
\begin{widetext}
\begin{equation}
Z=\int\displaystyle\prod_{k<\Lambda} d\phi_{k}e^{-S} =\int\displaystyle\prod_{k<\Lambda'} d\phi^{-}_{k}e^{-S_{0}[\phi_{k}^{-}]}\int\displaystyle\prod_{\Lambda' < k < \Lambda} d\phi_{k}^{+}e^{-S_{0}[\phi_{k}^{+}]}e^{-S_{I}[\phi_{k}^{-},\phi_{k}^{+}]}
\end{equation}
\end{widetext}
and this can be written using the over-arching Wilsonian effective action $S'[\phi^{-}]$
\begin{equation}
Z=\int \mathcal{D}\phi^{-}e^{-S'[\phi^{-}]}
\end{equation}
The functional form for this action will be the same with the parameters however "flowing" after the appropriate momentum re-scalings are being introduced. The procedures are totally standard and I do not see any reason to repeat them. We obtain 
\begin{equation}
S_{\zeta}[\phi']=\int d^{d}x[\frac{1}{2}\nabla \phi' \cdot \nabla \phi' + \frac{1}{2}\mu^{2}(\zeta)\phi'^{2}+g(\zeta)\phi'^{4}+...]
\end{equation}
where $\zeta$ is the rescaling factor $x'=\frac{x}{\zeta}$ and $\phi'(x')=\sqrt{\gamma '}\phi^{-}(x)$.
We generally remain with terms of the form 
\begin{equation}
e^{-S'[\phi^{-}]}=e^{-S_{0}[\phi_{k}^{-}]}\int\mathcal{D}\phi_{k}^{+}e^{-S_{0}[\phi_{k}^{+}]}e^{-S_{I}[\phi_{k}^{-},\phi_{k}^{+}]}
\end{equation}
There are well known perturbative methods of calculating the expectation value of the interacting term and to analyse the results in terms of Feynman diagrams probing the narrow area around a minimum that can be covered by a perturbative approach. Taking the trace over inaccessible modes reminds us of the various methods calculating the entanglement entropy of a black hole. The procedure is in many ways analogue. The coupling terms can be analysed from a non-perturbative point of view as well. Moreover, one has to take into account several phenomena linked to quantum entanglement of the various modes in this context. How does this work? There is of course the approach I took in ref. [1] where I discussed the homological algebraic terms and the alteration of the composition operation of the renormalisation semi-group in the context of imposing topological invariance as suggested by string theoretical T-duality. However, going back to low energies while keeping T-duality manifest proved to be quite complicated. Therefore, I here I take a different approach, linking the low energy scale of the standard model physics to the higher (but not string-theory high) energy scale we may encounter beyond the applicability region of the standard model. 
How can vacuum entanglement be described in terms of quantum fields? Considering a system of coupled harmonic oscillators in their ground state the vacuum is simply the unique state with zero excitations in any normal mode of the system. Considering such global modes as reference the vacuum is clearly non-entangled (fully separable). However, we can describe the same state in the basis of number states of each individual oscillator. With respect to such local modes the ground state is entangled. This is a generic feature of entanglement: the same state may be thought of as entangled if represented in some local basis or separable when one uses an overall global representation. The Minkowski vacuum state of a free quantum field is entangled with respect to the state space of any local observable although it may be separable with respect to Minkowski plane wave modes with zero particle in every mode. Excluding global bases due to horizons, be it black hole horizons or Rindler horizons leads to entanglement of various modes of our fields. 
However, in both cases entanglement allows us to introduce topologically non-trivial structures to operate in the context of ER-EPR duality. A similar way of thinking can be employed in the context of the renormalisation group approach. The cut-off is finally an artefact of our theory. While we are able to average out and calculate perturbatively the effects of most of the couplings between high and low energy modes, in some cases such an approach simply fails, and this may be because of the fact that the separation of the theory into modes localised in the momentum space requires the implementation of entanglement. 
To be more explicit consider the free scalar field in Minkowski spacetime and decompose it into plane wave modes. 
\begin{equation}
\mathcal{H}_{field}=\bigotimes_{k}L^{2}(\mathbb{R})_{k}
\end{equation}
the term $L^{2}(\mathbb{R})_{k}$ is simply the countably infinite harmonic oscillator state space corresponding to mode $k$. With this decomposition the Minkowski vacuum $\ket{0}$ is not entangled at all
\begin{equation}
\ket{0}_{M}=\bigotimes_{k}\ket{0}_{k}
\end{equation}
Now let us decompose the field into a left and a right component, for example according to the left and right half of the Rindler wedges
\begin{equation}
\mathcal{H}_{field}=\mathcal{H}_{left}\otimes \mathcal{H}_{right}
\end{equation}
What we obtain is a Minkowski vacuum which is a tensor product of two-mode squeezed states in pairs of Rindler modes 
\begin{equation}
\ket{0}_{M}=\bigotimes_{\omega}\ket{TMS}_{(\omega,1),(\omega,2)}
\end{equation}
and this shows clearly a bipartite entanglement between the left-right cut. We can imagine the quantum field as the continuum limit of a set of coupled quantum oscillators. This would show the image of a quantum field having a state space decomposable into local Hilbert spaces corresponding to every point in the space. Continuous matrix product states (cMPS) or continuous multiscale entanglement renormalisation ansatze (cMERA) allows such an expression to be constructed easily. It seems clear now that introducing a cut-off and separating the modes in high and low energy modes induces an entanglement which adds to the structure of $S_{I}[\phi_{k}^{+}, \phi_{k}^{-}]$. The question would be how? 
Let us perform the standard calculation of the RG procedure a bit further in the context shown above. We need to perform a functional integral $\int\mathcal{D}\phi_{k}^{\phi}$. This would amount to an expectation value of $e^{-S_{I}[\phi_{k}^{+}, \phi_{k}^{-}]}$ to be included in the final expression
\begin{equation}
e^{-S'[\phi^{-}]}=e^{-S_{0}[\phi_{k}^{-}]}\Bracket{e^{-S_{I}[\phi_{k}^{+}, \phi_{k}^{-}]}}_{+}
\end{equation}
But such a term will result in correlations across the cut-off region via topological terms visible in the action. The existence of such instanton-type contributions I showed in [7] although without introducing the current approach for quantum modes with a cut-off horizon. If we demand now high energy topology invariance it is clear that terms corresponding to low energy modes that become highly sensitive to the cut-off parameters will be strongly correlated with high energy modes which add correction terms. We will obtain therefore 
\begin{equation}
e^{-S'[\phi^{-}]}=e^{-S_{0}[\phi_{k}^{-}]}\Bracket{e^{-S_{I}[\phi_{k}^{+}, \phi_{k}^{-}]}}_{+}\Bracket{e^{-S_{Ent}[\phi_{k}^{+}, \phi_{k}^{-}]}}_{-}
\end{equation}
Each mode correction arising from the normal averaged term will require a low energy topological correction that has to take into account a high energy mode. Of course in the case of fermions this correction will overall average to zero leaving the fundamental scalars to be the sole receivers of such mode separation corrections.
 
\section{conclusion}
The separation of modes into high and low energy triggers a non-trivial entanglement which induces topological terms in the low energy region. These have to be compensated due to the high energy topology invariance requirement induced by T-duality. The result will be that there must appear low energy topological effects that would shift the coupling and mass of scalar fields downwards, potentially contributing at a better understanding of the naturalness problem beyond supersymmetry.

\end{document}